\documentclass[12pt]{iopart}
\usepackage{bbm, dsfont}
\usepackage{graphicx, graphics}
\usepackage{iopams}
\usepackage{float}
\expandafter\let\csname equation*\endcsname\relax
\expandafter\let\csname endequation*\endcsname\relax
\usepackage{amsmath,amsfonts,amsthm}
\usepackage{tikz-cd} 

\newcommand\norm[1]{\left\lVert#1\right\rVert}
\usepackage{hyperref}

\begin{document}

\title[Inferring Markovian quantum master equations]{Inferring Markovian quantum master equations of few-body observables in interacting spin chains}
%

\author{Francesco Carnazza$^1$, 
        Federico Carollo$^1$,
        Dominik Zietlow$^2$,
        Sabine Andergassen$^1$,
        Georg Martius$^2$ and
        Igor Lesanovsky$^{1,3,4}$}
\address{$^1$ Institut für Theoretische Physik and Center for Quantum Science,
        Universität Tübingen, 
        Auf der Morgenstelle 14, 
        72076 Tübingen, Germany}
\address{$^2$ Max Planck Institute for Intelligent Systems,
              Max-Planck-Ring 4, 72076 Tübingen, Germany}
\address{$^3$ School of Physics and Astronomy, University of                Nottingham, Nottingham, NG7 2RD, UK}
\address{$^4$ Centre for the Mathematics and Theoretical 
            Physics of Quantum Non-Equilibrium Systems,
            University of Nottingham, Nottingham, NG7 2RD, UK}

\begin{abstract}
Full information about a many-body quantum system is usually out-of-reach due to the exponential growth  --- with the size of the system --- of the number of parameters needed to encode its state. Nonetheless, in order to understand the complex phenomenology that can be observed in these systems, it is often sufficient to consider dynamical or stationary properties of local observables or, at most, of few-body correlation functions. These quantities are typically studied by singling out a specific subsystem of interest and regarding the remainder of the many-body system as an effective bath. In the simplest scenario, the subsystem dynamics, which is in fact an open quantum dynamics, can be approximated through Markovian quantum master equations.
Here, we formulate the problem of finding the generator of the subsystem dynamics as a variational problem, which we solve using the standard toolbox of machine learning for optimization. 
This dynamical or ``Lindblad" generator provides the relevant dynamical parameters for the subsystem of interest. 
Importantly, the algorithm we develop is constructed such that the learned generator implements a physically consistent open quantum time-evolution. We exploit this  to learn the generator of the dynamics of a  subsystem of a many-body system subject to a unitary quantum dynamics. We explore the capability of our method to recover the time-evolution of a two-body subsystem and exploit the physical consistency of the generator to make predictions on the stationary state of the subsystem dynamics. 
\end{abstract}

\section{Introduction}

Artificial neural network methods have established themselves as a versatile tool in many areas of physics  \cite{CarleoSc,Mehta2019,ML1}. Just to mention few examples, their application ranges from the classification of phases of matter \cite{Carrasquilla2017} over scattering and reflectivity analysis \cite{Greco2019,Greco_2021} all the way to the learning of topological states \cite{Deng2017}.
Moreover, neural networks have been effectively employed to encode the quantum state of both closed \cite{CarleoSc,Valenti2021} and open \cite{Carleo_Hartmann,nagy2019,reh2021timedependent,Yoshioka2019} quantum systems. 
          \begin{figure}[!t]
\centering
\resizebox{\textwidth }{!}{\includegraphics{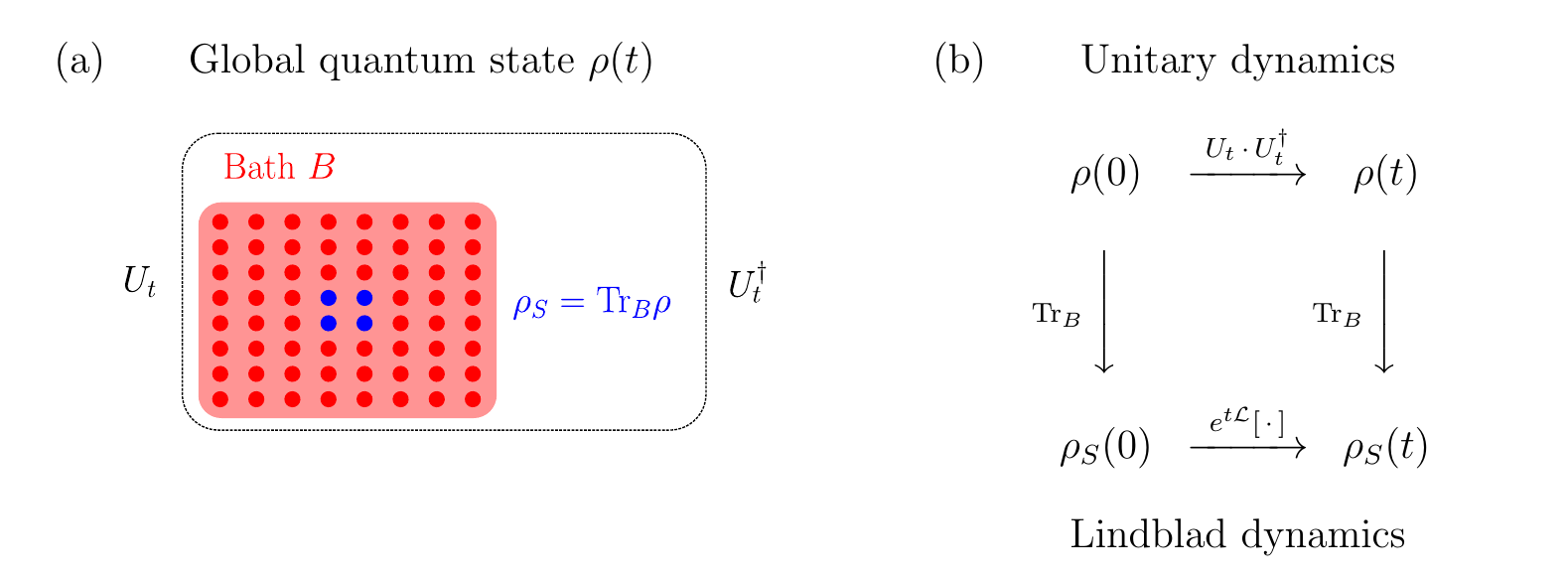}}

                \caption{ \textbf{Effective Lindblad generator for  the dynamics of a subsystem embedded in a many-body quantum system.}
                The full quantum state $\rho(t)$ of a many-body system subject to a unitary quantum dynamics evolves  according to the unitary (Schrödinger) operator $U_t$, as $\rho(t)=U_t\rho(0) U_t^\dagger$. (a) Properties about a subsystem $S$ of such a  many-body system are described by the so-called reduced quantum state $\rho_S$, which is  obtained by tracing out the remainder of the system (or, as we call it here, the bath), $\textbf{$\rho_S(t)={\rm Tr}_B\rho(t)$}$. (b) The time-evolution of the reduced subsystem state $\rho_S(t)$ can thus be obtained by tracing out the bath in the evolved many-body quantum state: $\rho_S(t)={\rm Tr}_B\rho(t)$ (see rightmost arrow in the diagram). If the emergent subsystem dynamics is Markovian, the time-evolution of any initial state of the subsystem, $\rho_S(0)$, can be formulated as an open quantum dynamics, characterised by a time-independent generator $\mathcal{L}$, which effectively accounts for the presence of the bath (see bottom arrow in the  diagram).}
        \label{fig:bath_system}
    \end{figure}
Recently, machine learning methods have also been applied to obtain effective dynamical generators for local degrees of freedom of a many-body quantum system \cite{Mazza2020}. The idea is the following: let us assume that one is interested in the time-evolution of a subsystem $S$ of a larger quantum system, as depicted in the sketch in Fig.~\ref{fig:bath_system}(a). For convenience, we refer to the complement of $S$, i.e.~the remainder of the quantum system, as the ``bath'' denoted by $B$. The full information about properties of $S$ is contained in the so-called reduced quantum state $\rho_{S}$, obtained by ``integrating out'' the bath $B$. Assuming that the full quantum system is subject to a Hamiltonian quantum dynamics implemented by the unitary operator $U_t$, this simply means that the time-evolved reduced quantum state of $S$ is given by $\rho_S (t) = \mathrm{Tr}_B(U_t\rho U_t^\dagger)$ [see Fig.~\ref{fig:bath_system}(b)], where $\mathrm{Tr}_B$ indicates the trace over the degrees of freedom of $B$. Under certain physical conditions, e.g.~weak $S$-$B$ coupling and very large bath $B$, such a subsystem time-evolution is well captured by Markovian open quantum dynamics \cite{breuer2002theory, Gorini1976, lindblad1976generators}. Formally, this entails the existence of a time-independent generator $\mathcal{L}$ of a dissipative quantum dynamics such that [see also Fig.~\ref{fig:bath_system}(b)]
\begin{equation}\label{master}
\rho_S(t) = \text{Tr}_B \left( U_t \rho(0) U_t^\dagger \right) \cong
e^{t\mathcal{L}}\left[\rho_S(0)\right]\, ,
\end{equation}
where $\rho(0)=\rho_S(0)\otimes \rho_B(0)$ and $\rho_{S/B}(0)$ is the initial state of the subsystem/bath, respectively. Interestingly, by exploiting simple neural network architectures, it was shown that such an approximate generator $\mathcal{L}$ can be found also beyond the typically considered settings, and that it can satisfactorily describe the subsystem dynamics whenever non-Markovian effects are negligible \cite{Mazza2020}. These findings are relevant since they allow one to infer, from the time-evolution of local degrees of freedom, the physical processes underlying the dynamics of quantum (sub)systems. This can, for instance, deliver insight into the dynamical effects behind an order-parameter time-evolution in non-equilibrium phase transitions or shed light on relaxation and thermalization effects in closed systems \cite{ETHreview}. Moreover, as we show here, since the neural network provides the dynamical generator of a Markovian time-evolution, the latter can be used to directly target stationary properties of the subsystem. However, the generator $\mathcal{L}$ learned by the neural network architecture of Ref.~\cite{Mazza2020} is not guaranteed to implement a valid, i.e. physical, dynamics (see below for details). This is a problem since it can lead to physically inconsistent predictions on the dynamics or on stationary properties of the subsystem state  \cite{Deen1971,Byrd2003,Kimura2003}.

In this paper we use a matrix parametrization of the Lindblad generator, and optimise it using data generated with exact numerical simulations. 
One can see this as an encoding of the matrix elements of the generator into the variational parameters of a simple neural network [see Fig.~\ref{fig:network} below]. This interpretation allows us to ``learn" the optimal variational parameters by  means of the standard tools of machine learning. For actual implementation and optimization of the network, we used the automatic differentiation toolbox Pytorch \cite{PyTorch}. 
Importantly, our parametrization is such that the learned dynamical generator $\mathcal{L}$ is constrained to be  physically consistent. 
Such a generator is meant to approximate the dynamics of the subsystem state $\rho_S(t)$ in the sense of Eq.~\eqref{master}. While the neural network, which is  agnostic to the fact that the dynamics of $\rho_S(t)$ is a result of tracing out the degrees of freedom of the bath $B$, always retrieves a time-independent generator $\mathcal{L}$. How well such an approximation can  reproduce the dynamics of the subsystem $S$ [see Eq.~\eqref{master}] clearly depends on the dynamical regime considered, e.g.~strong or weak interactions, determining whether the subsystem dynamics is effectively Markovian or not. Here, we show how the learned generator $\mathcal{L}$ can be used both to extrapolate the subsystem dynamics to times which have not been used to train the network and to predict stationary state properties of the subsystem. We illustrate our ideas by applying our neural network architecture to the reduced dynamics of a two-spin subsystem embedded in a larger quantum spin chain. 
        
\section{Formulation of the problem}

We consider a many-body quantum system partitioned into a subsystem $S$ and the remainder $B$, which in our setting plays the role of a bath. 

The separation implies that the full Hilbert space $\mathsf{H}$ is obtained as the tensor product $\mathsf{H}=\mathsf{H}_S\otimes\mathsf{H}_B$ of the Hilbert space of the subsystem ($\mathsf{H}_S$) and of the bath ($\mathsf{H}_B$). Here, we work under the assumption that $\mathsf{H}$ is finite-dimensional with dimension $m$. For instance, for the spin-$1/2$ models considered later [see Fig.~\ref{fig:models} and Eqs.~\eqref{Hpbc}-\eqref{Hobc}] $m=2^N$, where $N$ is the total number of spins. 
The quantum state of the full system is described by means of a density matrix $\rho(t)$, which must be positive semi-definite and must have trace equal to one in order to comply with the probabilistic interpretation of quantum mechanics. As such, the space of all possible states of the many-body system is the (convex) subspace of all positive semi-definite unit-trace matrices, $\mathsf{S(H)}\subset\mathbb{M}(m)$, where $\mathbb{M}(m)$ is the algebra of square matrices of dimensions $m$. Under the assumptions that the many-body system is subject to a unitary dynamics implemented by the full system Hamiltonian $H_{S+B}$, its state at time $t$ is given by $\rho(t)=U_t\rho(0) U_t^\dagger$, where $U_t=e^{-it H_{S+B}}$. 
    
The full information about the time-evolution of the degrees of freedom belonging to subsystem $S$ is contained in the reduced density matrix $\rho_S(t)$. As mentioned in the introduction, in certain settings the dynamics of the subsystem state can be approximated  through a Markovian open quantum dynamics implemented by a so-called Lindblad generator $\mathcal{L}$ \cite{Gorini1976,lindblad1976generators}, see Eq.~\eqref{master}. In these cases, $\rho_S(t)$ would effectively obey a quantum master equation 
    \begin{equation}\label{gks_rho}
    \frac{d \rho_S}{dt} =\mathcal{L}[\rho_S] = \mathcal{H}[\rho_S]+\mathcal{D}[\rho_S],
    \end{equation}
where we have dropped the explicit time dependence. The map $\mathcal{H}$ accounts for the Hamiltonian contributions to the dynamics while dissipative effects, uniquely associated with the interaction between $S$ and $B$, are encoded in $\mathcal{D}$. The general form of these terms (for finite-dimensional systems) is
    \begin{equation}\label{gks}
     \mathcal{H}[\cdot]= -i[H,\cdot],\qquad
     \mathcal{D}[\cdot]= \frac{1}{2}\sum_{i,j=1}^{d^2-1} c_{ij} \left( [F_i,\cdot F_j] + [F_i\cdot, F_j]\right)\, .
    \end{equation}
The operator $H$ is the Hermitian Hamiltonian of the subsystem only and describes its quantum coherent evolution. The operators $F_i$, with $i=1,2,\dots d^2$ and $d$ being the dimension of the subsystem Hilbert space, form an orthonormal basis of the subsystem algebra of operators $\mathbb{M}(d)$. Here, without loss of generality, we choose this basis with the property that all its elements  are Hermitian $F_i=F_i^\dagger$ and that $F_{d^2}$ is proportional to the identity, $F_{d^2}=\frac{1}{\sqrt{d}}\mathbbm{1}$. (Note that this latter term is not included in the double sum appearing in $\mathcal{D}$.) The orthonormality condition thus reads $\text{Tr}(F_i F_j)=\delta_{ij}$, where $\delta_{ij}$ is the Kronecker delta, showing that all elements $F_i$ but $F_{d^2}$ are traceless. 
The matrix $c$, with elements $c_{ij}$ is called the Kossakowski matrix, and it describes dissipative effects on $S$ due to its interaction with $B$. Such a matrix must be positive semi-definite in order for the dynamical map implemented by $\mathcal{L}$, $\mathcal{T}_t = e^{\mathcal{L}t}$, to be completely positive \cite{Gorini1976,lindblad1976generators};
we recall here that a map $ \mathcal{T}:\mathbb{M}(d)\rightarrow \mathbb{M}(d)$ is said to be  completely positive if, for any dimension $k$, the map
$\mathcal{T}\otimes \mathbbm{1}_k: \mathbb{M}(d) \otimes \mathbb{M}(k) \rightarrow \mathbb{M}(d)\otimes \mathbb{M}(k)$ is also positive, that is, it preserves the semi-positivity of the matrices it acts upon. 
Since $c$ is semidefinite positive, it can be diagonalized by means of an unitary transformation $h$: $h^\dagger c h= {\rm diag}(\gamma_1,...,\gamma_{d^2-1})$, with non-negative eigenvalues $\gamma_i$. The Lindblad generator can then be represented in its ``diagonal" form, via the operators $J_i=\sum_{i=1}^{d^2-1}h_{ij}F_{j}$ Eq.~\eqref{gks}, 
\begin{equation}
    \frac{d \rho_S}{dt} = -i [H,\rho] + \sum_{i=1}^{d^2-1}\gamma_i
    \left(
    J_i \rho J_i^\dagger
    -\frac{1}{2}
    \{
    J_i^\dagger J_i, \rho
    \}
    \right),
    \label{Lind_diag}
\end{equation}
where the operators $J_i$ are called jump operators.
The problem we address in this work is that of learning --- by means of neural network methods --- a time-independent physical dynamical generator $\mathcal{L}$ for the dynamics of a subsystem embedded in a unitarily evolving many-body system (see Fig.~\ref{fig:bath_system}). We are moreover interested in investigating in which parameter regimes the dynamics implemented by such a generator satisfactorily reproduces the time-evolution of the subsystem. From a technical perspective, it is convenient to pass from a representation of $\mathcal{L}$ as a map acting on the density matrix $\rho_S$ to that of a matrix acting on a vector representation of $\rho_S$ itself. We therefore discuss in the following how the information contained in $\rho_S$ can be encoded in a vector using the coherence-vector formalism \cite{Byrd2003}, and subsequently derive the ensuing matrix-representation of the Lindblad generator $\mathcal{L}$. We note that while there are several ways of such vectorizing a the density matrix, the one adopted here allows us to rewrite the full quantum dynamical problem in terms of real numbers only (see below), which is convenient for the implementation of neural network algorithms.  

Employing the (Hermitian) orthonormal basis of $\mathbb{M}(d)$ introduced above, we can write any density matrix $\rho_S\in \mathbb{M}(d)$ as a linear combination of the $F_i$. The coherence vector $\textbf{v}=(v_1,v_2,...,v_{d^2-1},1\sqrt{d})$ is then the vector in $\mathbbm{R}^{d^2}$ that gathers the corresponding expansion coefficients: 
    \begin{equation}
    \rho_S = \frac{\mathbbm{1}}{d}
           + \sum_{i=1}^{d^2-1} F_i v_i.
           \label{eq:coh-vec}
    \end{equation}
We note that the trace-normalization condition, $\text{Tr}(\rho_S)=1$, implies $[\textbf{v}]_{d^2}=1/\sqrt{d}$ which has been taken out of the sum. Furthermore, we note that $\textbf{v}$ is a real vector since both $\rho_S$ and the $F_i$ are Hermitian. Equation \eqref{eq:coh-vec} shows the one-to-one correspondence between $\rho_S$ and the coherence vector $\textbf{v}$, which we can exploit to conveniently represent the subsystem's density matrix. To make this idea more concrete, we discuss the example of spin-$1/2$ particle, whose Hilbert space dimension is $d=2$. In this case, the basis $\{F_i\}_{i=1}^4$ can be chosen to be proportional to the Pauli matrices $\{\sigma^x/\sqrt{2},\sigma^y/\sqrt{2},\sigma^z/\sqrt{2},\mathbbm{1}/\sqrt{2}\}$ with 
    \begin{equation}
    \sigma_x \equiv
    \sigma_1 =
    \begin{pmatrix}
    0 & 1 \\
    1 & 0 \\
    \end{pmatrix},
    \qquad
    \sigma_y \equiv
    \sigma_2 =
    \begin{pmatrix}
    0 & -i \\
    i & 0 \\
    \end{pmatrix},
    \qquad
    \sigma_z \equiv
    \sigma_3 =
    \begin{pmatrix}
    1 & 0 \\
    0 & -1 \\
    \end{pmatrix}
    \end{equation}
    such that the density matrix takes the form
    \begin{equation}
    \rho_S = \frac{\mathbbm{1}}{2} + \frac{\sigma_1 v_1 + \sigma_2 v_2 +\sigma_3 v_3 }{\sqrt{2}}\, .
    \end{equation}
The coherence vector parametrization is then $\textbf{v}=(v_1,v_2,v_3,1/\sqrt{2})$, which is reminiscent of the usual Bloch vector (a part for the last unimportant element and the different normalization of the Pauli matrices). To obtain the coherence vector elements from the density matrix one computes the traces
\begin{equation}\label{traces}
    v_i = \text{Tr}(F_i \rho_S)  \equiv \langle F_i\rangle \, ;
\end{equation}
as highlighted in the above equation, knowing the coherence vector basically amounts to knowing the expectation value of all possible observables of the subsystem.  

We now need to understand how the quantum master equation \eqref{gks_rho} translates into an evolution equation for the elements of the coherence vector. This can be done by taking a generic density matrix $\rho_S$ as in Eq.~\eqref{eq:coh-vec}, computing the action of the generator on it, $\mathcal{L}[\rho_S]$, and expanding this matrix into the basis formed by the operators $F_i$. This procedure, detailed in \ref{L_star}, provides the matrix representation $\textbf{L}$ of the map $\mathcal{L}$ which evolves the coherence vector $\textbf{v}$ through the equation
    \begin{equation}
        \frac{d\textbf{v}(t)}{dt}=
        \textbf{L} \textbf{v}(t), \qquad 
        \textbf{L} \equiv \textbf{H} + \textbf{D}\, . \label{eqn:Ldecomp}
    \end{equation}
Here, analogously to what done for $\mathcal{L}$, we have decomposed the matrix representation of the generator $\textbf{L}$ into a coherent Hamiltonian part, $\textbf{H}$, and a dissipative one, $\textbf{D}$. We note that, since the generator $\mathcal{L}$ is Hermiticity-preserving, the matrices $\textbf{H}$ and $\textbf{D}$  are real-valued. 
In particular, as shown in \ref{L_star}, the matrix $\textbf{H}$ is given by  
    \begin{equation}\label{eq:H*}
    \begin{split}
        &\textbf{H}_{ij} = 
        -4\sum_{k=1}^{d^2-1} f_{ijk} \boldsymbol{\omega}_k,
        \quad i,j \in \{1,2..,d^2-1\} ,\\ 
         &\textbf{H}_{id^2} = 
         \textbf{H}_{d^2i} = 0,
         \quad i \in \{1,2..,d^2\}.
    \end{split}
    \end{equation}
    where $\boldsymbol{\omega}$ defines the expansion of the Hamiltonian 
    matrix $H$ in Eq.~\eqref{gks} over  the orthonormal set 
    $\{F_i\}_{i=1}^{d^2}$:  $H = \sum_{i=1}^{d^2-1} \boldsymbol{\omega}_i F_i$. 
    The so-called antisymmetric structure constants $f_{ijk}$ are defined as
    $f_{ijk}    = -\frac{i}{4} \text{Tr} ([F_i,F_j]  F _k )$.
    The dissipative part is instead given by (see detailed derivation in \ref{L_star})
    \begin{equation} \label{eq:D*}
    \begin{split}
        &\textbf{D}_{mn} =
        -8\sum_{i,j,k=1}^{d^2-1} 
        \left( f_{mik} f_{njk} \mathrm{Re}(c)_{ij}  + 
          f_{mik} d_{njk} \mathrm{Im}(c)_{ij} \right), 
          \quad m, n \in \{1,2..,d^2-1\} ,\\
        & \textbf{D}_{md^2}=-4\sum_{i,j=1}^{d^2-1} f_{imj}\text{Im}(c)_{ij},
        \qquad \textbf{D}_{d^2m}=0,
        \quad m \in \{1,2..,d^2\},
    \end{split}
\end{equation}
where the so-called symmetric structure constants $d_{ijk}$ are given by
$d_{ijk}   = \frac{1}{4} \text{Tr}(\{F_i,F_j\}    F_k )$, and the matrix $c$ is the Kossakowski matrix appearing in Eq.~\eqref{gks}.

As we show below, by exploiting a neural network we can learn the matrix representation $\textbf{L}$, which propagates the degrees of freedom of a subsystem of a many-body system undergoing unitary quantum dynamics. The learned Lindblad generator $\mathcal{L}$ will by construction be physically valid, i.e. it will implement a completely positive and trace-preserving dynamics. However --- while it is always possible to find $\textbf{L}$ --- how well the learned subsystem dynamics reproduces the exact one depends on how much the latter can be actually approximated by a Markovian open quantum dynamics. 

\section{The architecture and the training procedure}
\begin{figure}
\centering
\resizebox{0.5\textwidth }{!}{\includegraphics{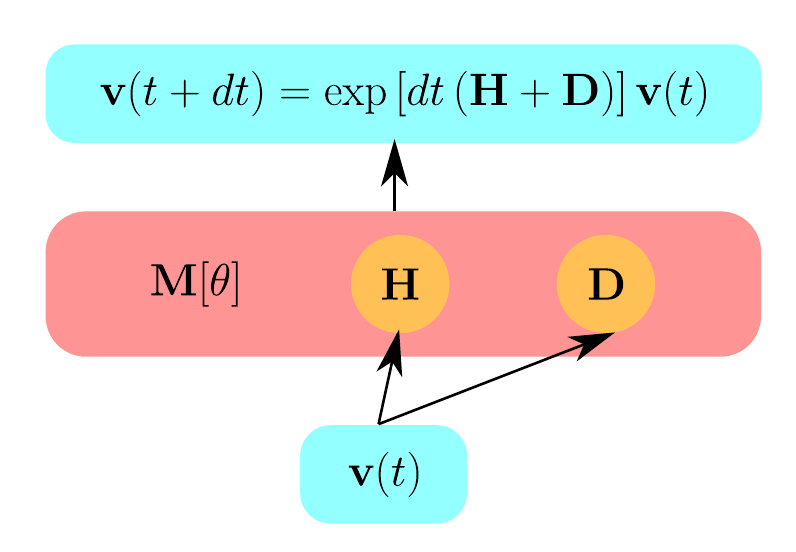}}
\caption{
\textbf{Architecture of the interpretable artificial neural network layer.} The network that we exploit takes as an input the coherence vector at time $t$, $\textbf{v}(t)$, and outputs its value at the next discrete time step: $\textbf{v}(t+dt) = \textbf{M}[\theta]\textbf{v}(t)$. Here $\textbf{M}[\theta]=e^{dt(\textbf{H}+\textbf{D})}$. The goal is to find the optimal parameters $\theta$ defining the matrices $\textbf{H}$ and $\textbf{D}$ by training the network with exact simulation data on the input and output coherence vector. The network is parametrized in a way that the learned generator is automatically guaranteed to be physically consistent.
The elements of the matrices $\textbf{H}$ and $\textbf{D}$ are functions of the variational parameters --- the weights of the network --- $\theta =\{\boldsymbol{\omega},X,Y\}$
through equations $\eqref{eq:H*}$ and $\eqref{eq:D*}$ and must be optimized according to the loss function,  Eq.~\eqref{cost}. 
}
\label{fig:network}
\end{figure}

In the following we discuss how the matrix $\textbf{L}$ is learned by a neural network through training it with data obtained from simulating the exact unitary many-body quantum dynamics, see sketch in Fig.~\ref{fig:network}. According to Eq.~\eqref{eqn:Ldecomp} we have the decomposition $\textbf{L} = \textbf{H}+\textbf{D}$. Here, $\textbf{H}$ is defined in terms of the $d^2-1$ unconstrained real parameters $\boldsymbol{\omega}$ appearing in  Eq.~\eqref{eq:H*}. The matrix $\textbf{D}$ is parametrised by the complex Hermitian matrix $c$ [see  Eq.~\eqref{eq:D*}], which is constrained to be positive semi-definite. To enforce this constraint by construction, we express $c$ as $c=\textrm{Z}^\dagger \textrm{Z}$ for some complex 
matrix $\textrm{Z}= X + i Y$, where $X$ and $Y$ are real matrices. The parameters of our neural network which are to be found via training are thus $\theta = \{\boldsymbol{\omega}, X, Y\}$, and our parametrization
automatically ensures the complete positivity of the dynamics generated by $\textbf{L}$. 
Note that, since we want to learn a Markovian dynamics, we assume the parameters $\theta     = \{\boldsymbol{\omega}, X, Y\}$ to be time independent.
Moreover, only the symmetric real matrix $\text{Re}(c) = X^tX +Y^tY$ and the skew symmetric real matrix $\text{Im}(c) = X^tY - Y^tX$ appear in Eq.~\eqref{eq:D*} [see also Eqs.~\eqref{Dmn} and \eqref{Dmd2}], showing that the network indeed only makes use of real parameters.

Given an input coherence vector  at time $t$, $\mathbf{v}_\textrm{in} = \mathbf{v}(t)$, the network $\mathbf{M}$ (see Fig.~\ref{fig:network}), which is a function of parameters $\theta$, is trained to output $\mathbf{v}_\textrm{out} = \mathbf{M}[\theta]  \mathbf{v}_\textrm{in}$ such that $\mathbf{v}_\textrm{out}$ is  close to $\mathbf{v}(t+dt)$, that is the coherence vector after a discrete time step of length $dt$. In this way, the network learns how to propagate the coherence vector by an infinitesimal time-step and this information fully specifies the matrix $\textbf{L}$ [cf.~Eq.~\eqref{eqn:Ldecomp}].

Training data consist of a set of trajectories of the time-evolution of $\textbf{v}$, starting from an initial state $\textbf{v}_0$
up to a fixed time $T$ 
\footnote{More specifically, the training data is formed by 50 exactly evolved trajectories, each with the initial conditions as discussed in subsection (\ref{subs:init_cond}).
The batch size is 256 with 512 batches per epoch, over 20 epochs.  
Of the generated data,  $80\%$ is used as training set, while $20 \%$ is employed as validation, to check that the model does not over fit.}.
Each trajectory is thus a list of snapshots $(\textbf{v}_0, \textbf{v}(dt), \textbf{v}(2dt), ... , \textbf{v}(T) )$. These trajectories are obtained by exactly evolving the full many-body quantum state $\rho$ under the action of the global Hamiltonian $H_{S+B}$, tracing out the degrees of freedom of the bath at each time step, and computing $\textbf{v}$ as in Eq.~\eqref{traces}. For each choice of parameters in the Hamiltonian, $n$ trajectories were used as training data $D$. The network parameters $\theta$ are optimised through the Adam optimization algorithm \footnote{
Hyperparameters used in Adam were
learning rate  $\textrm{Alpha} = 10^{-3}$, 
$\textrm{Beta}_1= 0.9, \textrm{Beta}_2=0.999$ and $\textrm{Epsilon}=10^{-8}$.
} 
\cite{kingma2017adam} to minimize the common mean squared error cost function 
\begin{equation}\label{cost}
    l(\theta) =
    \mathbb{E}_{
    \textbf{v}(t),\textbf{v}(t+dt)
    \sim D} [
    \lVert \mathbf{M[\theta]
    } \mathbf{v}(t)-  \mathbf{v}(t+dt)\rVert  ^2].
\end{equation}
Here, the expectation $\mathbb{E}$ is taken over the training dataset $D$.

To summarise, the parameters of the network are $ \theta = \{\boldsymbol{\omega}, X, Y\}$, where $\boldsymbol{\omega}$ is a real vector defining $\textbf{H}(\boldsymbol{\omega})$ and $X,Y$ are real matrices defining $\textbf{D}(X,Y)$. 
In particular, $ \theta = \{\boldsymbol{\omega}, X, Y\}$ provides the ``weights" of the network.
One needs first to use the matrices $X$ and $Y$ to compute the semidefinite positive matrix $c = (X-iY)^t(X+iY)$.
After a choice of the basis for the Hilbert space of the subsystem has been made,
one can compute the structure constants and use them to map the variational parameters to the matrices $\textbf{H}$ and $\textbf{D}$
via Eqs.~\eqref{eq:H*} and ~\eqref{eq:D*}.
Then $\textbf{H}(\boldsymbol{\omega})$ and $\textbf{D}(X,Y)$ are applied to $\textbf{v}_{\textrm{in}}$ as
 \begin{equation}
    \textbf{M}[\theta]\textbf{v}_{\textrm{in}} 
    \equiv 
    e^{dt(\textbf{H}(\boldsymbol{\omega})+\textbf{D}(X,Y))}\textbf{v}_{\textrm{in}}
\end{equation}
and the variational parameters are optimized by minimizing the loss function in Eq. ~\eqref{cost}.
Here $dt$ is the time step used in the exact integration of the dynamics upon which the network is trained. 
In general, in order to learn all the relevant dynamical features of the subsystems one should make sure that $dt$ is sufficiently small, so that all timescales are captured. We have considered $dt=0.01/\Omega$, as  we observed that with such a time step the details of the exact evolution were 
well reproduced. Since creating the artificial dataset though exact diagonalization was
the most time consuming task. Therefore, it not efficient to use a too small $dt$.
Once the network has been trained, it is possible to retrieve the Hamiltonian part $\textbf{H}$ and the Kossakowski matrix $c$, we give a concrete example of this in \ref{Kossa_H}, where we also give a brief explanation of their forms.

\section{Many-body models and subsystem of interest}

\label{section:Mod}
\begin{figure}
\centering
	\resizebox{0.65\textwidth }{!}{\includegraphics{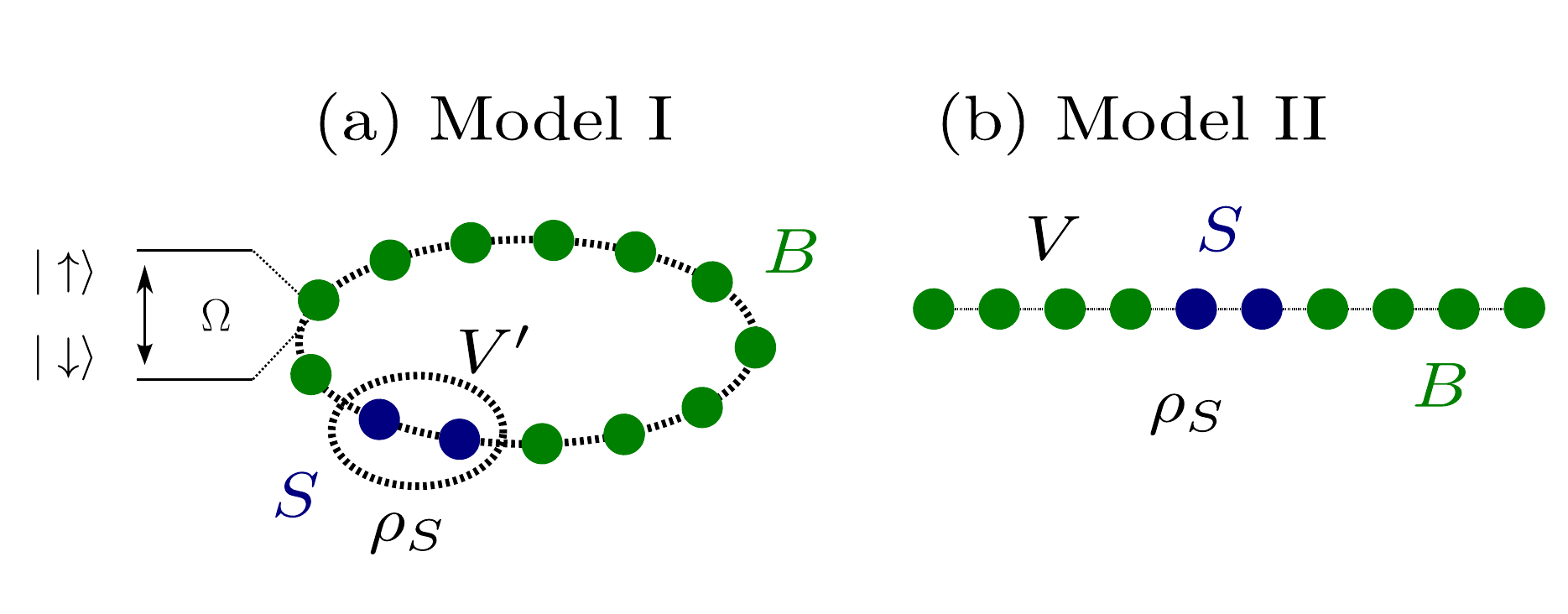}}
        \caption{ 
        \textbf{Spin-chain models.}
        We exploit our neural network architecture to learn the reduced dynamics of a two-spin subsystem (highlighted in the panels) embedded into two different many-body quantum spin chains, one with closed boundary conditions (Model I) and one with open boundaries (Model II). In both settings, the global time-evolution is governed by a Hamiltonian,  $H_\textrm{I}$ [cf.~Eq.~\eqref{Hpbc}] and $H_\textrm{II}$ [cf.~Eq.~\eqref{Hobc}], respectively. (a) Quantum spin chain with closed boundary conditions. In this case, the Hamiltonian
        $H_\textrm{I}$ contains only nearest-neighbour interactions. As free parameters for the model, we consider the  inverse temperature $\beta$ for the initial state of the bath spins, and the interaction strength of the subsystem within itself and with the bath $V'$. These parameters are varied and the error $I_\textrm{Err}$, between the exact dynamics and the (Markovian) one predicted by the neural network model is quantified through
        Eq.~\eqref{eq:error}.
        (b) Quantum spin chain with open boundary conditions. The full system Hamiltonian $H_\textrm{II}$ features long-range interactions, 
        which decay with a power-law with exponent $\alpha$. This exponent and the coupling strength between the spins $V$ are varied in order to investigate the quality of the approximation of the dynamics obtained from the generator calculated by the network. The subsystem of interest contains two contiguous spins located at the centre of a chain 
        of (even) length $N$.
        }
\label{fig:models}
\end{figure}

For testing our ideas, we use the above-discussed neural network architecture to learn the generator of the reduced dynamics for two neighbouring spins embedded in a one-dimensional chain.
The spin chain is thus partitioned into two parts as shown in Fig.~\ref{fig:models}. The first one is formed by two nearest-neighbouring sites which for convenience are labelled as spin $1$ and spin $2$.
This part acts as the ``subsystem of interest''.
The other part is the remainder of the lattice and is regarded as the bath.
We assume that a Hamiltonian $H_{S+B}$ governs the dynamics of the full system quantum state $\rho(t)$. Training data is computed by evolving this state according
to the unitary evolution operator $U_t = e^{-itH_{S+B}}$ and tracing out the degrees of freedom of the bath.
As initial state, $\rho_0=\rho(t=0)$, of the unitary many-body dynamics we consider product states of the form
\begin{equation}
\rho_0 = \rho_S(0) \otimes \rho_B(0).
\end{equation}
The reduced density matrix at time $t$ is given by $\rho_S(t) =\text{Tr}_B U_t \rho_0 U^\dagger_t$. 

We consider two different different quantum spin chain models, as shown in Fig.~\ref{fig:models}, and refer to them as model I and II.  Model I has closed boundaries, i.e. the spins form a ring, while model II has open boundaries. For Model I, the full subsystem-bath dynamics is governed by a nearest-neighbour interaction Hamiltonian
\begin{equation}
\label{Hpbc}
H_\textrm{I} = \frac{\Omega}{2} \sum_{i=1}^{N} \sigma_i^x +
      V \sum_{i=3}^{N-1} n_i n_{i+1} +
      V'\left( n_L n_1 + n_1 n_2 + n_2 n_3 \right).
\end{equation}
The operator $n_i=\frac{1+\sigma^z_i}{2}$ denotes the projector onto the spin up state of the $i$-th spin. The constants $V$ and $V'$ are the strength of the interaction among neighbouring spins: $V$ is associated with interactions among the bath spins, while $V'$ is associated with interactions between the subsystem sites and between the subsystem and the bath sites next to it [see also Fig.~\ref{fig:models}(a)]. The contribution proportional to 
$\Omega$, which is the same for all sites, parameterises a transverse ``laser'' field and drives Rabi oscillations between single spin states. The motivation behind studying the Hamiltonian $H_\textrm{I}$ is that the first and the second spin define the subsystem and their interaction with the bath is assumed to be controllable.

Model II, which features open boundary conditions, is governed by a Hamiltonian that features power-law interactions:
\begin{equation}
\label{Hobc}
H_\textrm{II} = \frac{\Omega}{2} \sum_{i=1}^N\sigma^x_i + V\sum_{i<j}\frac{n_i n_j}{|i-j|^\alpha},
\end{equation}
where $\alpha \geq 0$ defines the power with which the interactions decay over distance. Note, that unlike in Model I, there is no specific choice for the coupling constant among bath and system spins.
In \ref{Kossa_H} we adopt Hamiltonian II to give an example of the form of the retrieved Hamiltonian part and Kossakowski matrix of the Lindblad generator.
There, we deem the restriction of Eq.~\eqref{Hobc} to a system of two spins as $H_\textrm{II}^{(2)}$, and undertake a brief study of how the bath affects the form of this Hamiltonian.
Hamiltonians (\ref{Hpbc}) and (\ref{Hobc}) are variants of the Ising model with transverse and longitudinal field as well as to the so-called PXP model \cite{sun2008,Ates2012}. Experimentally they can be realized, for example, with Rydberg atoms \cite{Bloch2012, kim2018,Ebaldi2021,Browaeys2020,Bloch_rev}. 

Given that we focus on a system whose reduced density matrix is that of two spins, the dimension $d$ of the reduced Hilbert space $\mathsf{H}_S$ is four, and a natural choice for the basis $\{F_i\}_{i=1}^{d^2}$ is given by the two-spin Pauli group generators 
$\{
 \sigma^x/\sqrt{2}, \sigma^y/\sqrt{2}, \sigma^z/\sqrt{2},
 \mathbbm{1}_2/\sqrt{2}
 \}^{\otimes 2}$. Hence, $\{F_1,F_2, ... ,F_{d^2}\}$ = $\{\mathbbm{1}_2\otimes \sigma^x/2, \mathbbm{1}_2\otimes \sigma^y/2,...,\mathbbm{1}_4/2\}$.
If $N$ is the length of the spin chain, the subsystem of interest is identified by the spins at positions $1$ and $2$, such that each element in $\textbf{v}$ corresponds to either a two-body expectation value $\langle  \sigma_{m}^1 \sigma_n^{2} \rangle$,  or to a single-body expectation value 
$\langle \sigma^1_m\rangle, (\langle \sigma^{2}_m\rangle)$, with
$m,n\in \{1,2,3\}$, and 
\begin{equation}
\sigma^k_l =  \mathbbm{1}_2^ {\otimes k-1} \otimes \sigma_l \otimes \mathbbm{1}_2^ {\otimes(N- k)}.
\end{equation}

\subsection{The initial conditions}\label{subs:init_cond}
For both models, the initial conditions of the system are chosen as the product of a thermal state for the bath 
$\rho_{\textrm{Bath}} \propto e^{-\beta H_{\mathrm{I/II}}}$, with $\beta$ the inverse temperature, and  a valid density matrix  $\rho_{S}$ for the system $S$.
Model II was studied to  investigate whether and how the algorithm is able to capture the dynamics in the presence of longer-ranged interactions. To be able to focus on this aspect, we decided to take a fixed initial state for the bath (without changing the temperature) and, for simplicity, we considered the infinite-temperature ones ($\beta =0$).
For the study of both models we choose the initial density matrix $\rho_S$ in the following way:
we consider two random $d\times d$ real matrices $M$ and $N$ whose entries are taken from a Gaussian 
distribution centred at zero and with standard deviation one and we construct the density matrix according to
\begin{equation}\label{eq:rho_0}
\rho_S(0)=\frac{(M+iN)^\dagger (M+iN)}{\text{Tr}\left[(M+iN)^\dagger (M+iN)\right] }\, .
\end{equation}
This ensures  that $\rho_S(0)$ is a 
Hermitian positive semi-definite unit-trace matrix.

\section{Results and discussion}
In the standard treatment of open quantum systems, the dynamics of the reduced state of a subsystem  can be approximated by means of Markovian open quantum dynamics only when certain conditions are met \cite{breuer2002theory} (see e.g.~Ref.~\cite{ciccarello2017} and references therein for a different derivation of a Markovian quantum  master equations). These include, for instance, a weak subsystem-bath coupling, an infinitely large bath with a continuous dispersion relation and a large separation of time-scales between the subsystem and the bath dynamics. 
    
For the Models I and II [see Eq.~\eqref{Hpbc} and \eqref{Hobc}], not all of the above conditions are met. For instance, while it is possible to tune the parameters in a way that the interaction between $S$ and $B$ is weak, the bath $B$ will always be, in our setting, a finite-dimensional object whose Hamiltonian possesses a discrete spectrum. 
Nonetheless, given that our network is capable of retrieving a time-independent generator for the subsystem from training data, it is natural to ask whether the dynamics implemented by such a generator can provide an approximate description of the subsystem state also in the considered settings.

\begin{figure}[b!]
\centering
	\resizebox{1.0\textwidth }{!}{\includegraphics{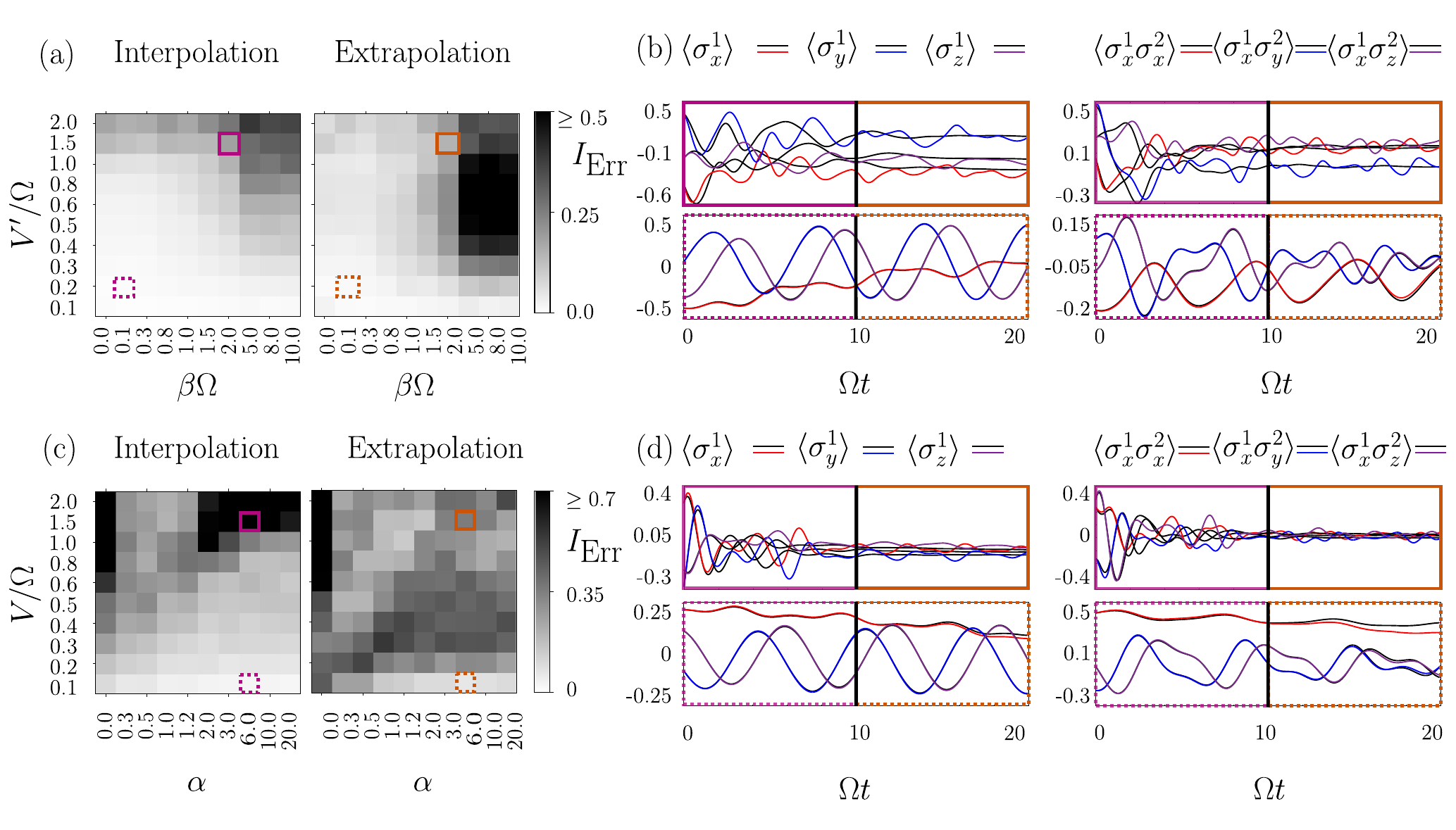}}
        \caption{
        \textbf{Interpolation and extrapolation error.} (a) Scan of the error, $I_\textrm{Err}$, defined by Eq.~\eqref{eq:error} for different values of the parameters $\beta$ and $V'$ for Model I. The considered spin-chain length is $N=11$. The interpolation error $I_\textrm{Err}$ is computed from $T_\textrm{in}=0$ to $T_\textrm{fin}=10/\Omega$, where $T_\textrm{fin}$ is the time up to which the network is trained on the exact 
        trajectories. The extrapolation error $I_\textrm{Err}$ refers to the region in time $t$ where the network is used to extrapolate from $T_\textrm{in}= 10/\Omega$ up to unseen times $t\le T_\textrm{fin}$. We choose $T_\textrm{fin} = 20 /\Omega$. The discrete time step used in the calculations is $dt=0.01/\Omega$. (b) Examples of how the network model performs, for Model I, in predicting the time-evolution of expectation values of quantum observables. We study both single-site (left panels) and two-site (right panels) spin observables. The top panels show a case taken from the parameter region characterised by large errors $I_\textrm{Err}$, while the bottom ones show a case taken from a region in which the error is small [see squares in panel (a)]. (c) Scan of the error for different values of the parameters $\alpha$ and $V$ for Model II. The considered spin-chain length is $N=10$. (d) Same as in panel (b) but for Model II. In panels (b) and (d), the black lines are predictions from the network model, and coloured lines are showing exact simulation results obtained from the dynamics of the full many-body system state.
        }\label{fig:intra_extra}
        \end{figure}

\begin{figure}[h!]
    \centering
	\resizebox{\textwidth }{!}{\includegraphics{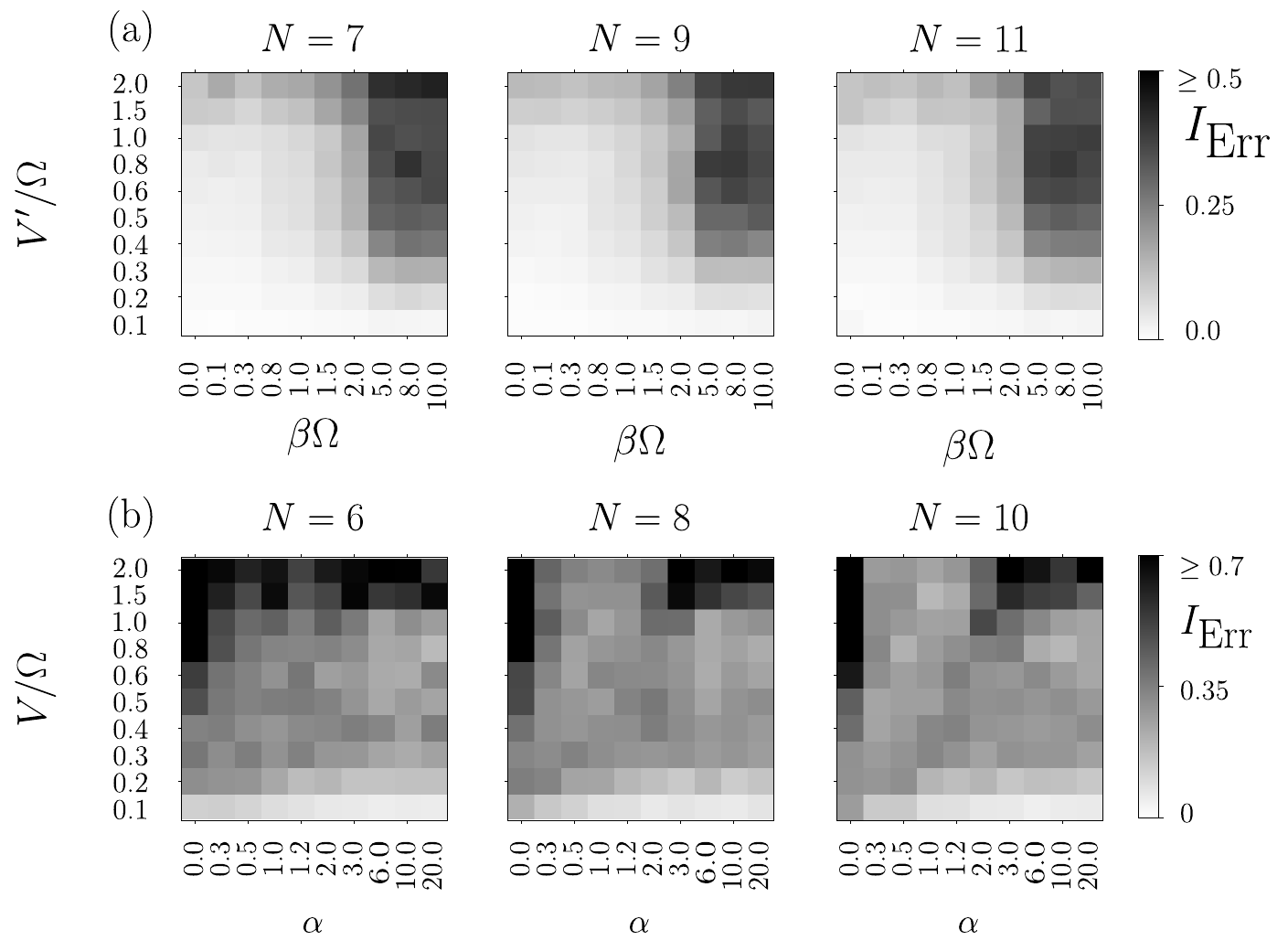}}
        \caption{ 
        \textbf{Scaling of the error with the bath size $N$ for Model I (a) and Model II (b).} (a) Scans of the error $I_\textrm{Err}$ for different values of the parameters $\beta,V'$ for Model I, produced for different choices of the chain length $N=7,9,11$.
        (b) Error $I_\textrm{Err}$ for Model II, for different choices of the parameters $ V,\alpha$ and different chain lengths $N=6,8,10$. For both models, the error is computed from time zero up to time $T_\textrm{fin} = 20/\Omega$. Here, in contrast to Fig.~\ref{fig:intra_extra}(a)-(c), no distinction is made between extrapolation and interpolation region. For Model II the error is the highest for the smallest of the system sizes investigated, as one may expect. No dependence on system size is discernible instead for model I.
}
\label{fig:scaling}
\end{figure}

To determine whether, and for which parameter range, a Markovian dynamical description is a valid one for the subsystem and, thus, whether the network can be used to predict the dynamical behaviour of its observables, we investigate the accuracy of the Markovian approximation obtained through the network by varying the parameters $(\beta, V')$ and $(\alpha,V)$ for model I and II, respectively.
To quantify the error made in the approximation, we define the error measure 
    \begin{equation}\label{eq:error}
    I_{\textrm{Err}}(T_\textrm{in},T_\textrm{fin}) =  \frac{1}{T_\textrm{fin}-T_\textrm{in}}\int _{T_\textrm{in}}^{T_\textrm{fin}} dt 
    \norm{ \rho_{\textrm{exact}}(t)- \rho_{\textrm{network}}(t)}_1.
    \end{equation}
Here, $\rho_{\textrm{exact}}$ is the (time-evolved) quantum state for $S$ obtained from the exact diagonalization of the full many-body problem, 
while $\rho_{\textrm{network}}$ is the subsystem dynamics as predicted by the network.
The trace norm $\norm{\sigma}_1$ for a $d\times d$ complex matrix $\sigma$, with eigenvalues 
    $\lambda_1,\dots,\lambda_d$ is given by
    \begin{equation}
            \norm{\sigma}_1 = \text{Tr}\sqrt{\sigma^\dagger \sigma} = \sum_{i=1}^{d} |\lambda_i| .
    \end{equation}
    In order to remove the dependence of the error from the specific initial condition considered, we consider the quantity $I_\textrm{Err}$, defined in Eq.~\eqref{eq:error}, averaged over 
    10 different trajectories, each with an independent initial condition (\ref{eq:rho_0}) [shown in Figs.~\ref{fig:intra_extra} and \ref{fig:scaling}].

\subsection{Subsystem dynamics}
We start our discussion by considering Model I, see Eq. (\ref{Hpbc}).
As apparent from Fig.~\ref{fig:intra_extra}(a),
for relatively low values of the coupling between the subsystem and the bath, the generator learned by the neural network can provide a good approximation of the reduced subsystem dynamics. This is also manifest from the bottom panels in Fig.~\ref{fig:intra_extra}(b); the time-evolution of the expectation values of subsystem observables, as predicted by the network, is in very good agreement with exact numerical data, both for single-site observables and for two-site ones. On the other hand, Fig.~\ref{fig:intra_extra}(a) clearly shows that the neural network approximation becomes less and less reliable upon increasing the coupling between the subsystem and the bath. We stress that this is not witnessing an issue occurring in the training of the neural network for these parameter regimes. The reason for this increase of the error is that, consistently with what expected form the weak-coupling approximation, for large enough coupling the reduced subsystem dynamics becomes non-Markovian and cannot be approximated by the time-independent generator learned by the network. 
We also note that a similar worsening of the Markovian approximation of the dynamics can be observed when lowering the temperature of the bath (increasing $\beta$), as becomes transparent from Fig.~\ref{fig:intra_extra}(a). 

As a first application of the learned generator of the subsystem dynamics, one can use it to extrapolate the dynamics of the subsystem observables to times which have not been seen during the training procedure. This is in particular true for the range of parameters for which the error in approximating the dynamics with the one retrieved by the neural network is relatively small, as shown in Fig.~\ref{fig:intra_extra}(b). The rationale is that whenever the subsystem dynamics is Markovian, the dynamical generator is time-independent and, once this is learned, it can also be used outside the training time-window. 

For Model II we find that when the interaction between sites decays fast enough ($\alpha \geq 1$) 
and when the coupling is low ($V/\Omega \approx 0.1$), the error $I_\textrm{Err}$ remains relatively small [see Fig.~\ref{fig:intra_extra}(c)]. In this regime, just as before, the learned generator can be used also for making predictions beyond training times, as shown in Fig.~\ref{fig:intra_extra}(d).
On the other hand, when the power $\alpha$ decreases, we observe an increase in the approximation error [see Fig.~\ref{fig:intra_extra}(c)]. This is due to the fact that the concomitant increase of the range of the interactions amounts to increase the interaction between $S$ and $B$. It is thus reasonable to expect that non-Markovian effects become more pronounced. 

In the following we investigate the behaviour of the error, for both models, when changing the length of the spin chain, i.e.~modifying the size of the bath, see Fig.~\ref{fig:scaling}. For Model I, we observe that the error remains very similar for the different system sizes explored. This may suggest that, in the case of nearest-neighbour interactions and for the model at hand, the considered sizes can be already considered sufficiently large for the remainder of the many-body system to act as a proper bath. For Model II, instead, there appears to be an improvement in the approximation upon increasing the size of the bath. This may be related to the fact that for long-range interactions, revivals in the subsystem dynamics --- due to the back-flow of information from the bath into the subsystem --- have stronger effects for smaller system sizes. 

For both models, our observations suggest that a Markovian approximation may be indeed justified, also for a bath formed by discrete degrees of freedom, provided that the latter is sufficiently large.
Moreover, we observe that for all parameter regimes, even in those in which the Markovian approximation is not quite satisfactory, the error tends to stay bounded. This means that the predicted long-time value of local observables seems to overall capture the actual behaviour of the subsystem, at least within the considered  time-window [cf.~Fig.~\ref{fig:intra_extra}(b-d)].
This seems to suggest that even for large errors in the interpolation, one may find actually a relatively small error in the extrapolation, see Fig.~\ref{fig:intra_extra}. This effect is seen in several parameter regimes, where the expectation values of subsystem observables is almost constant, but features small amplitude oscillations in the extrapolation window, as shown in Fig.~\ref{fig:intra_extra}. 

\begin{figure}
\centering
\resizebox{0.7\textwidth }{!}{\includegraphics{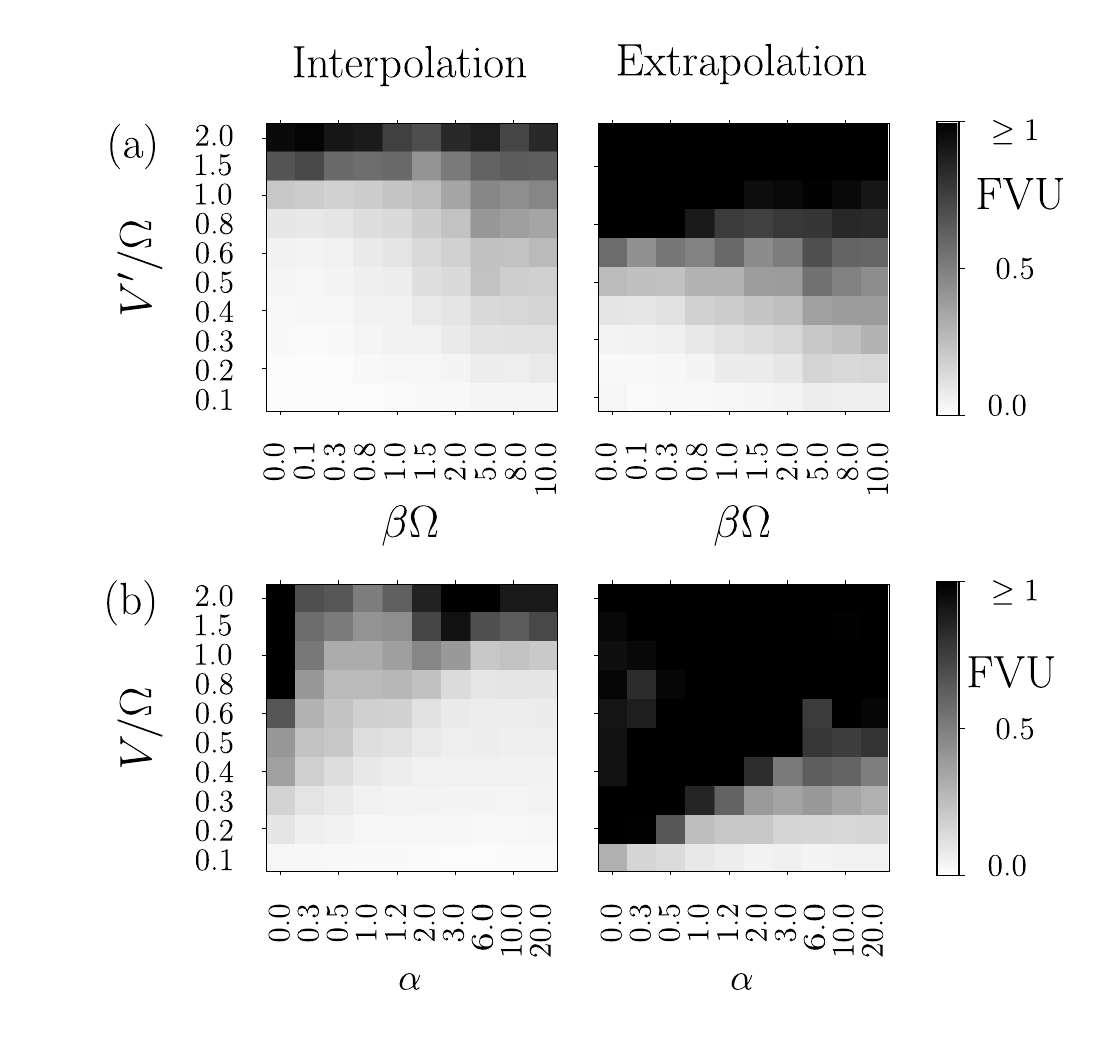}}
\caption{
\textbf{Fraction of variance unexplained (FVU)} (a) FVU of the interpolation and extrapolation error for Model I and a system size $N=11$ as a function of the system-bath interaction $V^\prime/\Omega$ and the inverse temperature $\beta\Omega$. (b) FVU of the interpolation and extrapolation error for Model II as function of the interactions strength $V/\Omega$ and the power-law exponent $\alpha$ of the interaction potential.}
\label{fig:FVU}
\end{figure}

In order to have a fairer comparison of the error between initial transient regime with large oscillations and the long-time one in which oscillations have very small amplitude, we introduce a relative error measure. We consider the so-called fraction of variance unexplained, or FVU. We compute this quantity between the coherence vector predicted by the network $v_i^\textrm{network}$ and the one retrieved from the exact simulation $v_i^\textrm{exact}$, as in the following equation 
\begin{equation}\label{FVU}
    \textrm{FVU} = \frac{1}{d^2-1}\sum_i^{d^2-1}
    \sqrt{
    \frac
    {\textrm{Var}(v_i^\textrm{exact} - v_i^\textrm{network})}
    {\textrm{Var}(v_i^\textrm{exact})}
    }.
\end{equation}
The set over which the variance Var is computed is given by the time snapshots with discrete time step $dt$, $v_i=(v_i(t_0),v_i(t_0+dt),...)$.
The fraction of variance unexplained normalises an error to the variance of the signal (here the exact coherence vector). 
The results for Model I and II are plotted in Fig.~\ref{fig:FVU} for system sizes $N=11$ and $N=10$, respectively, while results for smaller system sizes are reported in \ref{fvu}. 
Comparing Fig.~\ref{fig:FVU} with Fig.~\ref{fig:scaling}, we note that considering the FVU provides a smoother behaviour of the error. In particular, it allows to distinguish between the regions for which the error in the approximation $I_\textrm{Err}$ is small simply because the overall signal has small variance, from those where the error is small because the neural network dynamics correctly captures the behaviour of the subsystem.
The latter situation occurs for small values of the coupling strength in both Models I and II and for fast decaying interactions in model II [cf. Fig. \ref{fig:FVU}].
In this region the Markovian approximation is correct, and the network is able to reconstruct the dynamics almost exactly.

\subsection{Stationary behaviour}
\begin{figure}
\centering
	\resizebox{1.0\textwidth }{!}{\includegraphics{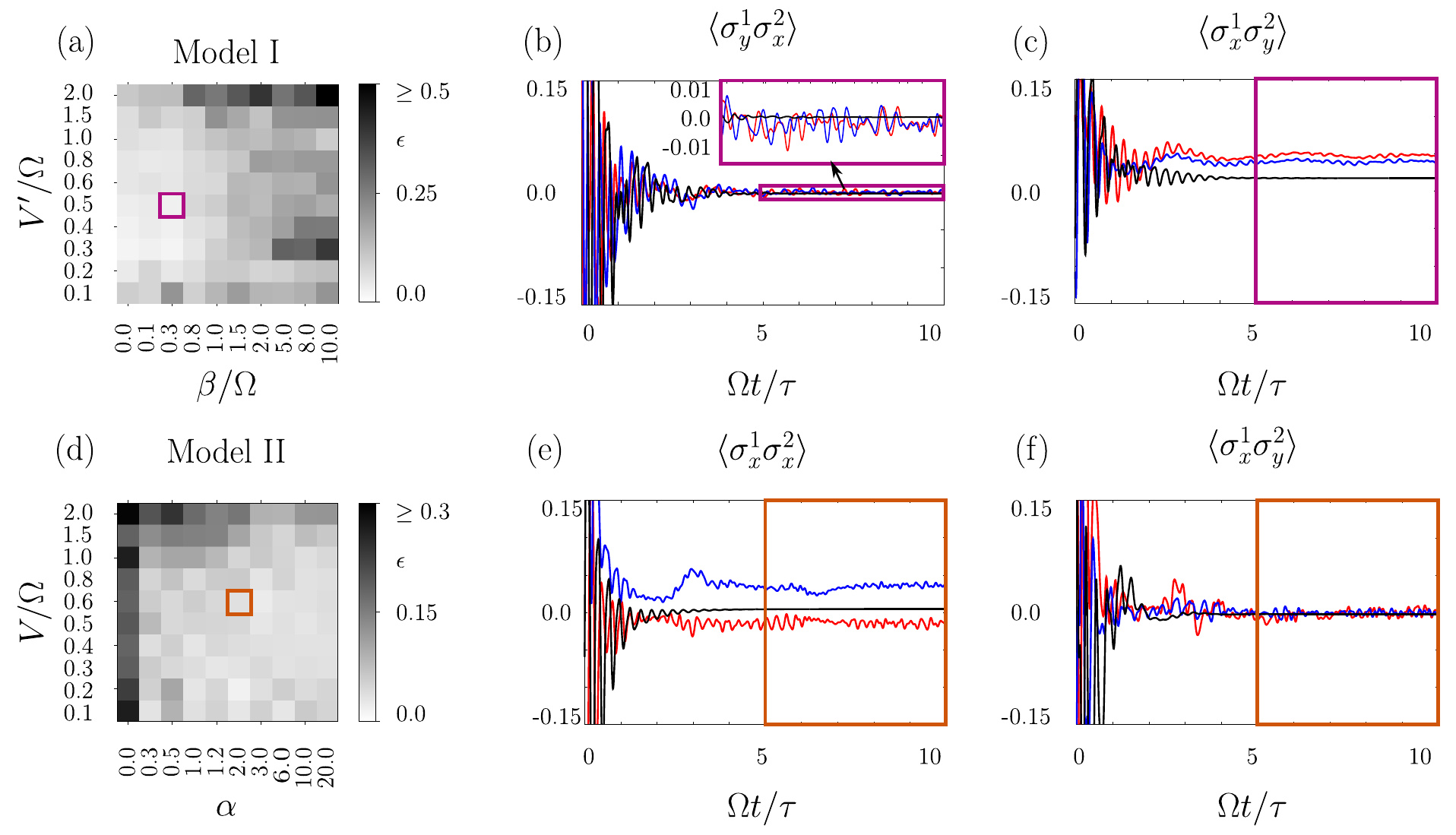}}
        \caption{
        \textbf{Stationary state dynamics.}
        In panels (a) and (d) the error $\epsilon$, defined by Eq.~\eqref{epsilon}, is depicted for Model I with $N=11$
        and Model II with $N=10$, respectively. The error is computed for different parameter combinations $(\beta,V')$ for Model I and $(V,\alpha)$ for Model II. For specific parameter choices the dynamics of some two-spin observables is shown in panels (b) and (c) for Model I and panels (e) and (f) for Model II. Red and blue lines indicate the same expectation value for two different initial conditions for the quantum state       $\rho_\textrm{exact}$. The black line indicates the evolution according to the learned $\textbf{L}$. The regions in time where the error $\epsilon$ is computed are marked by boxes.
        As the behaviour of single body observable does not differ much from the two-body  expectation values, we did  not report it.
        }\label{fig:stat}
        \end{figure}

Once the network has learned the matrix-representation $\textbf{L}$ of the Lindblad generator, it is also possible to investigate stationary properties of the system \cite{Vicentini2019,Carleo_Hartmann}. In our case, we can do this by studying the stationary-state coherence vector, which is nothing but the eigenvector $\textbf{v}_\textrm{st}$ associated with the zero eigenvalue of $\textbf{L}$:
$\textbf{L} \textbf{v}_\textrm{st}=0=\frac{d\textbf{v}_\textrm{st}}{dt}$. Such a vector provides the stationary density matrix of the subsystem $S$, see Eq.~\eqref{eq:st-ex}, which in our case is by construction ensured to describe a physically consistent quantum state. 
In principle, there is no reason to expect that the long time coherence vector $\textbf{v}_\textrm{exact}(t)$ will converge to $\textbf{v}_\textrm{st}$. This is because i) we are training the network for a finite time-window and ii) the bath is finite, and thus one would expect to observe, for long times, recurrence and revivals in the time-evolution of system observables. These effects are associated with a re-entering of the information scrambled from the subsystem into the bath into the subsystem again, and are associated with non-Markovian behaviour. Nonetheless, what we observe (shown in Fig.~\ref{fig:stat}), is that in some parameter regimes the agreement between $\textbf{v}_\textrm{st}$ and the long-time behaviour of the exact dynamics is remarkably good. 

To be more quantitative, let us define $E_\textrm{gap}$ as 
the smallest, in modulus, among the real parts of the non-zero eigenvalues of $\textbf{L}$ and let $\tau$ be $1/E_\textrm{gap}$. This latter quantity represents the time-scale of the approach to stationarity in a Markovian open quantum system. Moreover, in order to circumvent the issue that the exact subsystem dynamics always presents oscillations (also due to the finiteness of the bath), we define the following error measure
\begin{equation}\label{epsilon}
    \epsilon =  \norm{ \rho^*_{\textrm{exact}}- \rho_\textrm{st}}_1,
\end{equation}
where  
\begin{equation}
\rho_\textrm{st} = \frac{\mathbbm{1}}{d} + \sum_{i=1}^{d^2-1} [\textbf{v}_\textrm{st}]_iF_i\, ,\qquad 
    \rho^*_{\textrm{exact}} =  \frac{1}{(b-a)\tau}
    \int _{a\tau}^{b\tau} dt 
    \rho_{\textrm{exact}}(t)\, .
    \label{eq:st-ex}
\end{equation}
Thus, the error $\epsilon$ measures the distance between the stationary behaviour predicted from the network results and an averaged long-time behaviour of the exact solution. In particular, in order to be sure to address the stationary behaviour, at least in relation to what predicted by the network, we choose $a=5$, $b =10$. Within this time window --- provided that the exact dynamics is correctly captured by the time-independent generator --- the expectation value of local observables should have already converged to their stationary value. Given the finiteness of the bath, these will --- as mentioned above --- display residual oscillations around such an average stationary value which would be described indeed by $\rho^*_{\textrm{exact}}$. 
The discrete time step $dt$ for the exact integration and for the training of the network is chosen to be  $dt=0.01/\Omega$ as before. Once the training was complete, we could retrieve the Lindblad generator $\textbf{L}$ and its eigenvalues, and in particular $E_\textrm{gap}$ and the time scale $\tau=E_\textrm{gap}$. 
In Fig. \ref{fig:stat} the error in Eq.~(\ref{eq:st-ex}) is shown. It is computed by averaging over $10$ different initial conditions for $\rho_\textrm{exact}$. 
This error is rather different both from the error of Eq.~\eqref{eq:error} and more importantly from the FVU of Eq.~\eqref{FVU}. The
FVU is indeed used to compare the oscillations of the predicted and of the exact dynamics. However, by construction, the predicted dynamics is bound
to reach a stationary state, and the FVU will thus tend to be large. This signals that the learned dynamics is not able to reproduce the small oscillations around the stationary values.  Nonetheless, we can investigate how well the learned dynamics can capture the average value of this oscillations, which should provide their stationary value. To this end, a more appropriate error measure is the one given in Eq.~\eqref{epsilon}. 
One notices that while there are observables that approach the learned steady state $\textbf{v}_\textrm{st}$, e.g.\ $\sigma^1_y\sigma^2_x$ in model I, others do not. In certain cases, the exact time-evolution of the expectation of observables converges towards a long-time behaviour which depends on the initial conditions. This may be related to the fact that we are considering a finite bath. On the other hand, the learned dynamics always approaches the same stationary behaviour since, due to the existence of a finite gap $E_\textrm{gap}$, the steady state of the Lindblad generator is unique.

\section{Conclusions}
We have introduced a simple neural network whose parameters can be exactly mapped onto those of a Lindblad generator. Importantly, such a generator which is learned by the network from exact dynamical data is automatically ensured to be a physically consistent generator of a quantum Markovian dynamics. We have investigated the applicability of such an architecture to two different classes of spin models. Even though the considered physical settings are rather different from those known to give rise to Markovian subsystem dynamics, we find that, in certain parameter regimes, the network model provides a faithful approximation of the subsystem time-evolution. 

Future developments in the same spirit may include the adaptation to
architectures capable of encoding time correlations in  time series, such as long short term memory (LSTM) \cite{LSTM} or transformers \cite{Transformers}, which would allow for the learning of a time-dependent generator.
A different path to achieve time dependence would be that of learning the analytical solutions of the differential equations of motion \cite{SahooLampertMartius2018:EQLDiv}, or by numerically solving them by means of neural networks \cite{NODE}.

We have exploited the time-independent generator learned by the network in order to investigate stationary properties of the reduced subsystem state. This idea looks promising as a path towards capturing relevant long-time features such as thermalization effects \cite{ETHreview, Polkovnikov2011}. 
Finally, instead of learning the linear evolution of the density matrix, one may think of directly learning the evolution of an order parameter, such as the magnetization or particle density \cite{kharkov2021discovering}. This would entail machine learning of an effectively non-linear dynamical evolution within the state space of the order parameter. This directly leads to the question whether and how more involved neural network architectures permit an increased performance in determining effective generators and possibly allow an improved quantification of time dependence and non-Markovianity. 

The presented results open routes towards the understanding of complex non-equilibrium dynamics through a reduced number of (collective) degrees of freedom. This may ultimately allow to develop simplified descriptions of complex dynamical non-equilibrium processes which is not only of interest in fundamental research but may also be of importance when harnessing many-body phenomena in technological applications.

\section*{Acknowledgments}
The authors thank P. Mazza and M. Klopotek for valuable discussions. We acknowledge financial support from the Deutsche Forschungsgemeinschaft (DFG, German Research Foundation) under Germany’s Excellence Strategy – EXC-Number 2064/1 – Project number 390727645. IL acknowledges funding from the “Wissenschaftler Rückkehrprogramm GSO/CZS” of the Carl-Zeiss-Stiftung and the German Scholars Organization e.V., and through the DFG projects number 449905436. The authors thank the International Max Planck Research School for Intelligent Systems (IMPRS-IS) for supporting DZ.
\appendix

\section{Representation of the matrix $\textbf{L}$} \label{L_star}
In this appendix we derive equations (\ref{eq:H*}) and (\ref{eq:D*}) for the dynamics of  the coherence vector $\textbf{v}= (v_1,...,v_{d^2})$. The derivative of $v_h$ with respect to $t$ 
can be written as
    \begin{equation}
    \frac{d}{dt}v_h(t) =  \text{Tr}\left(F_h\frac{d}{dt}[\rho_S(t)]\right) 
    		           =  \text{Tr}\left(F_h\mathcal{L} [\rho_S(t)]\right) 
    		           = \text{Tr}\left(   \mathcal{L}^*[F_h ] \rho_S(t)\right),
    \end{equation}
    where in the first equality we used Eq.~\eqref{traces} and in the second one \eqref{gks_rho}. 
    The map $\mathcal{L}^*$ indicates the dual map of $\mathcal{L}$, i.e.~the one which evolves observables and not the subsystem state. The action of the superoperator $\mathcal{L}^*$ can be obtained from that of $\mathcal{L}$ via the cyclic property of the trace.
    Expanding 
    $\mathcal{L}^*[F_h] =\sum_{k=1}^{d^2}\text{Tr}(F_k \mathcal{L}^*[F_h])F_k$, one finds
    \begin{equation}\label{eq:coherence}
    \frac{d}{dt}v_h(t) = \sum_{k=1}^{d^2}\text{Tr}( \mathcal{L}^*[F_h]    F_k)  \text{Tr}(  F_k\rho_S(t)) = \left[\textbf{L}\textbf{v}(t)\right]_h,
    \end{equation}
    where we 
    the matrix $\textbf{L}$ is defined as 
    $\textbf{L}_{hk}=\text{Tr}( \mathcal{L}^*[F_h]    F_k) $. Explicitly, we obtain 
    \begin{equation}\label{dual}
                              \text{Tr}( \mathcal{L}^*[F_h]    F_k) =  \sum_{h=1}^{d^2}\text{Tr}\left( i [H,F_h] F_k  + \frac{1}{2} \sum_{i,j=1}^{d^2-1}c_{ij}( F^\dagger_j[F_h,F_i] F_k +  [F^\dagger_j,F_h]F_iF_k )\right).
\end{equation}
The comparison with Eq.~\eqref{gks} yields 
\begin{equation}
\textbf{H}_{kh} = i \text{Tr}([H,F_k] F_h), \qquad \textbf{D}_{kh} = \frac{1}{2} \sum_{i,j=1}^{d^2-1}c_{ij}\text{Tr} \left( F^\dagger_j[F_k,F_i] F_h +  [F^\dagger_j,F_k]F_iF_h \right).
\end{equation}
We can now introduce
the structure constants $d_{ijk}$ and $f_{ijk}$ for the basis $\{F_i\}_{i=1}^{d^2}$.
The structure constants characterise the commutation and anti-commutation relations of $\{F_i\}_{0\leq i < d^2}$ as
\begin{equation}
\{F_i,F_j\}  = 2\delta_{ij}\frac{\mathbbm{1}}{d}+\frac{1}{4}  \sum_{k=1}^{d^2-1}  d_{ijk}  F_k,\qquad
 [F_i,F_j]   = \frac{-i}{4} \sum_{k=1}^{d^2-1}  f_{ijk}   F _k.
\end{equation}
Since $\text{Tr}(F_iF_j) = \delta_{ij}$, one obtains
\begin{equation}
d_{ijk}   = \frac{1}{4} \text{Tr}(\{F_i,F_j\}    F_k ),\qquad
f_{ijk}    = -\frac{i}{4} \text{Tr} ([F_i,F_j]  F _k ).
\end{equation}
As a consequence, $d_{ijk}$ is fully 
anti-symmetric under exchange of two indices, while $f_{ijk}$ 
is fully symmetric.
Moreover, 
for self-adjoint $F_i = F^\dagger_i$,
the structure constants are real.

For the Hamiltonian part, we consider the expansion 
$H = \sum_{i=1}^{d^2-1}F_i\boldsymbol{\omega}_i$, with $\boldsymbol{\omega} =(\boldsymbol{\omega}_1,...,\boldsymbol{\omega}_{d^2-1})$
a $d^2-1$ dimensional real vector.
Then, the matrix elements of $\textbf{H}$ are determined by 

\begin{equation}\label{eq:H1}
           \textbf{H}_{mn} = i \sum_{k=1}^{d^2-1}\text{Tr}([F_k,F_m]F_n) \boldsymbol{\omega}_k.
\end{equation}
The Hamiltonian part can thus be rewritten as 
\begin{equation}\label{eq:H2}
\begin{split}
&\textbf{H}_{ij} = -4\sum_{k=1}^{d^2-1} f_{ijk} \boldsymbol{\omega}_k, 
\qquad  \textbf{H}_{id^2} = \textbf{H}_{d^2i} = 0, \\
\end{split}
\end{equation}
with $i,j \in \{1,2..,d^2-1\}$. Thus, the Hamiltonian part is skew-symmetric and can be 
parametrised by a single vector $\boldsymbol{\omega}$. 

The dissipative part instead takes the form
\begin{eqnarray} \label{Dmn}
\textbf{D}_{mn}
&=\frac{1}{2}\sum_{i,j}^{d^2-1}  c_{ij}  \text{Tr}\left([F_m,F_i] F_n F_j  +[ F_j ,F_m]F_i F_n \right) \nonumber \\
& =-8\sum_{i,j,k=1}^{d^2-1} ( f_{mik} f_{njk} \text{Re}(c)_{ij}  + f_{mik} d_{njk} \text{Im}(c)_{ij} )
\end{eqnarray}
for $1\leq m<d^2, 1 \leq n<d^2$.
In the last line we used the Hermiticity of the Kossakowski matrix  $c$, i.e. that its real (imaginary) 
part is (skew-) symmetric.
For  the matrix elements of $\textbf{D}$ with $1\leq m<d^2$, $n=d^2$  one has
\begin{equation} \label{Dmd2}
\begin{split}
\textbf{D}_{md^2} = 
\frac{1}{2}\sum_{i,j,k=1}^{d^2-1}  c_{ij} \text{Tr}\left( F_i [F_m,F_j] +[ F_i ,F_m]F_j \right)=
-4\sum_{i,j =1}^{d^2-1} f_{imj}\text{Im}(c)_{ij}, \\
\end{split}
\end{equation}
where we used the cyclic property of the trace in the second equality, and the fact that $f_{imj}$ and $\text{Re}(c)_{ij}$ 
are antisymmetric in the indices $(ij)$,
while $\text{Re}(c)_{ij}$ is symmetric. 
Finally, 
for $1\leq n<d^2$, one obtains $\textbf{D}_{d^2n}=0$.

\section{The fraction of variance unexplained (FVU)} \label{fvu}
\begin{figure}[H]
\centering
	\resizebox{0.8\textwidth }{!}{\includegraphics{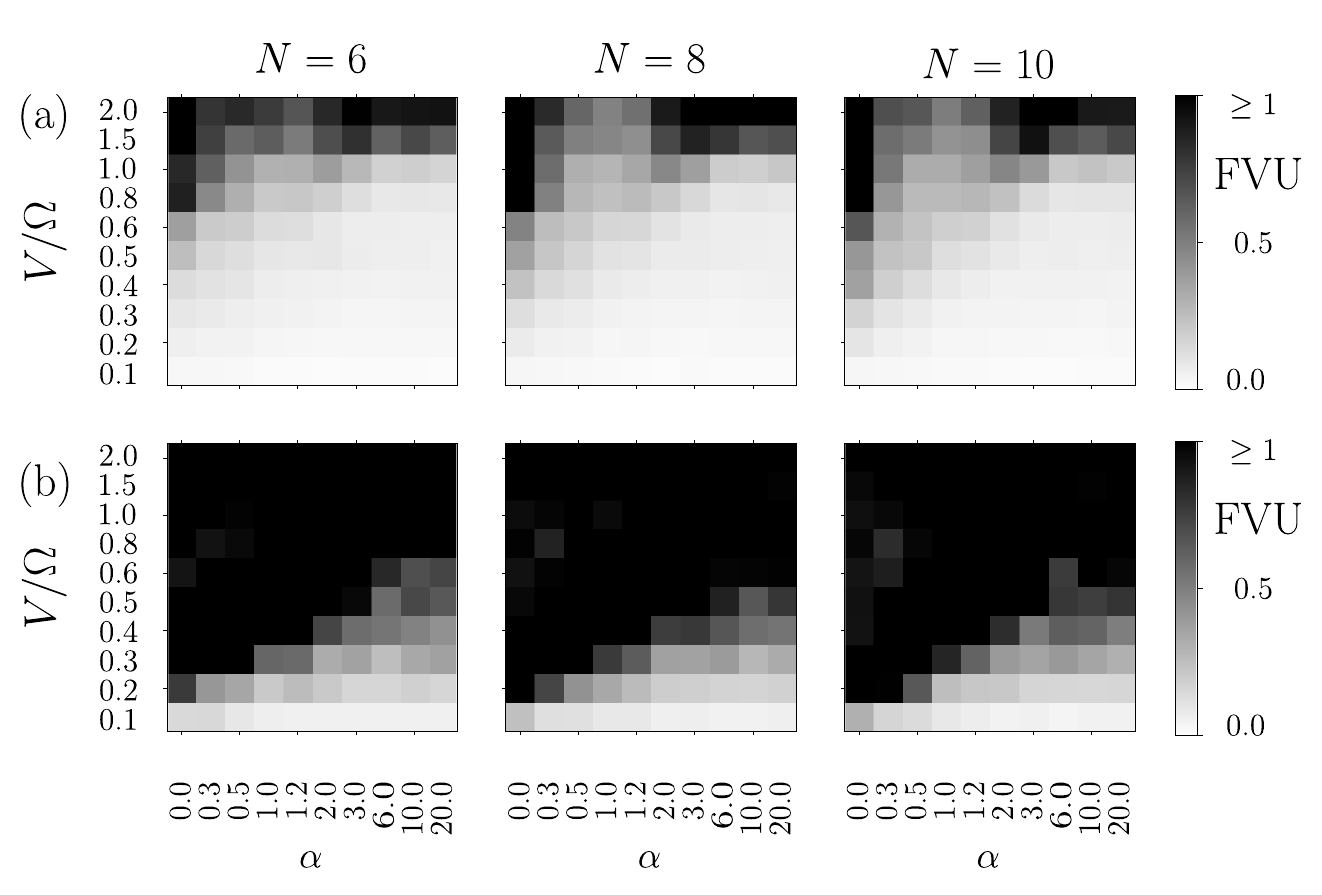}}
        \caption{
        \textbf{\text{FVU} in various spin chain lengths for Model I.} In the upper row (a) is reported the interpolation and in (b) the extrapolation region.
        }\label{fig:FVU_I}
        \end{figure}

\begin{figure}[H]
\centering
	\resizebox{0.8\textwidth }{!}{\includegraphics{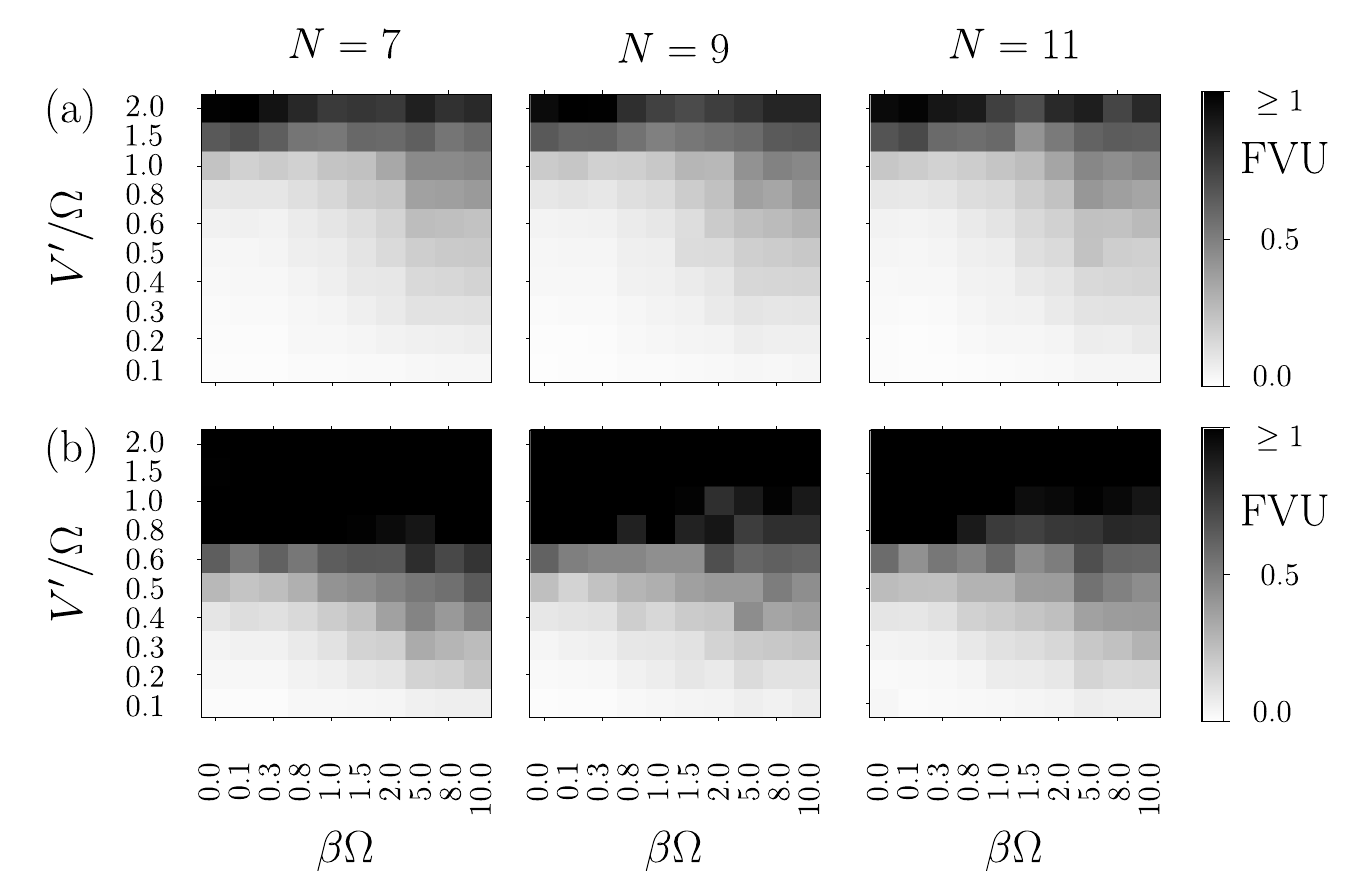}}
        \caption{
        \textbf{\text{FVU} in various spin chain lengths for Model II.} In rows (a) and (b) the FVU is shown in the interpolation and extrapolation regime respectively. 
         }\label{fig:FVU_II}
        \end{figure}

In this appendix we report additional results on the FVU for different system sizes $N$. In Fig.~\ref{fig:FVU_I} we show results for Model I both in the interpolation [see panel Fig.~\ref{fig:FVU_I}(a)] and in the extrapolation [see panel Fig.~\ref{fig:FVU_I}(b)] regimes. In Fig.~\ref{fig:FVU_II} we report analogous results for Model II.


\section{Interpretability of the Network}
\label{Kossa_H}
\begin{figure}[H]
        \centering
        \resizebox{0.8\textwidth }{!}{\includegraphics{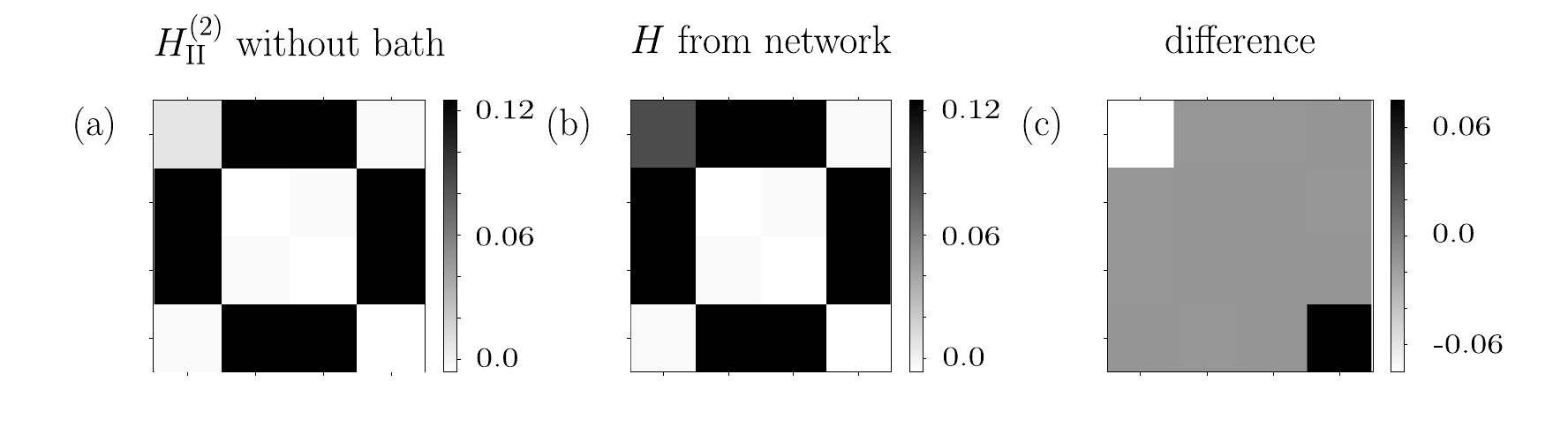}}
        \caption{
        \textbf{Two-spin Hamiltonian with and without bath. } 
        (a) Matrix elements of the Hamiltonian II for a system of two spins $  H_\textrm{II}^{(2)} $ 
        with parameters $V'/\Omega=0.1$ and $\alpha = 0.3$. In order to compare this with the learned Hamiltonian, we have removed the irrelevant contribution proportional to the identity, i.e.~we consider $H_\textrm{II}^{(2)}\to H_\textrm{II}^{(2)}-{\rm Tr}  \left(H_\textrm{II}^{(2)}\right)$. (b) Matrix elements of the  Hamiltonian $H$ learned by the generator for same parameters as in (a) and  a system of six spins. (c) Difference between the exact Hamiltonian in (a) and the learned one in (b), which shows the emergence of a term proportional to $\sigma_z \otimes \mathbbm{1} + \mathbbm{1} \otimes \sigma_z$. 
        }\label{fig:H_1_order}
\end{figure}

We here give an example of how one can extract information about the subsystem dynamics from the network
and interpret it as the Hamiltonian and dissipative part of the Lindblad equation.
In particular we focus here on model II for the parameters $V'/\Omega=0.1$
and $\alpha = 0.3$.
In Fig.~\ref{fig:H_1_order}(a) we plot the matrix elements of the Hamiltonian of model II for 
a chain of only two spins, already discussed below Eq.~\eqref{Hobc}.
In Fig.~\ref{fig:H_1_order}(b) is plotted the one retrieved from the network, that is, the Hamiltonian part $H$ of the Lindblad generator.
Their difference is given in ~Fig.~\ref{fig:H_1_order}(c). As shown, this difference is proportional to a term $\mathbbm{1}\otimes\sigma_z+\sigma_z \otimes \mathbbm{1}$.
 
This contribution can be explained by means of a ``mean-field" treatment of the Hamiltonian involving the density-density interactions between the bath and the subsystem.  Indeed, it results from taking terms like $\langle n_i\rangle n_j$, where $i$ is a site of the bath and $\langle n_i \rangle$ stands for expectation value,  while $j$ a site of the two-spin subsystems, once terms proportional to the identity are removed.

\begin{figure}[H]
        \centering
        \resizebox{0.6\textwidth }{!}{\includegraphics{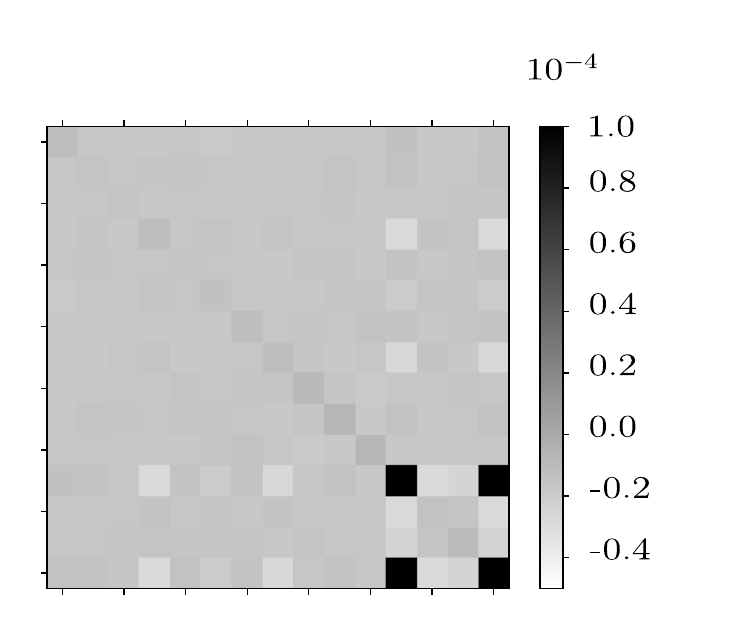}}
        \caption{
        \textbf{Real part of the Kossakowski matrix.} Real part of the Kossakowski 
        matrix retrieved from a six spin system evolved according to Hamiltonian II
        with parameters $V'/\Omega=0.1$ and $\alpha = 0.3$. This shows that the largest values are associated with combination of operators $\sigma_z \otimes \mathbbm{1}$ and $\mathbbm{1} \otimes \sigma_z$ with equal weights. This means that the relevant dissipative effect is a dephasing noise, associated with a jump operator $J\propto \sigma_z \otimes \mathbbm{1} + \mathbbm{1} \otimes \sigma_z$. Note that, in this case, the entries of the Kossakowski matrix are small. This is due to the fact that we are considering small values of $V'/\Omega$ and, in this regime, dissipation is expected to be of second-order in $V'/\Omega$.   
        }\label{fig:Kossa}
\end{figure}

For the same system it is possible to also retrieve the Kossakowski matrix.
We plot the real part of its entries in Fig.~\ref{fig:Kossa}. The basis of choice for the Hilbert space of subsystem is the same as in the main text, and it is given by $\{F_1,F_2, ... ,F_{d^2}\}$ = $\{\mathbbm{1}_2\otimes \sigma^x/2, \mathbbm{1}_2\otimes \sigma^y/2,...,\mathbbm{1}_4/2\}$.
This shows that the most relevant entries are associated with dephasing implemented by a ``collective" jump operator [cf.~Eq.~\eqref{Lind_diag}] of the form 
$J\propto \sigma_z \otimes \mathbbm{1} + \mathbbm{1} \otimes \sigma_z$.

\section*{References}
\bibliographystyle{iopart-num}
\bibliography{bib}

\end{document}